\documentstyle[amssymb,epsfig,12pt]{article}

\newcommand{\lsim}{\mathrel{\mathop{\kern 0pt \rlap
  {\raise.2ex\hbox{$<$}}}
  \lower.9ex\hbox{\kern-.190em $\sim$}}}
\newcommand{\gsim}{\mathrel{\mathop{\kern 0pt \rlap
  {\raise.2ex\hbox{$>$}}}
  \lower.9ex\hbox{\kern-.190em $\sim$}}}
\newcommand{\gagamma}{g_{a\gamma\gamma}}

\begin{document}
\title{Physics Potential and Prospects for the CUORICINO and CUORE experiments}%
\author{I. G. Irastorza$^{a}$, A. Morales$^{a}\footnote{corresponding
author: amorales@posta.unizar.es}$, S. Scopel$^{b}$, S. Cebri\'{a}n$^{a}$ \\ (from the CUORE Collaboration \footnote{University of California, Berkeley; University of Como/INFN; University of Firenze/INFN; L.N. del Gran Sasso; L.N. di Legnaro; Leiden University; University of Milano/INFN; University of South Carolina; University of Zaragoza} )}%
%\address{Laboratorio de F\'{\i}sica Nuclear y Altas Energ\'{\i}as. Universidad de Zaragoza. 50009 Zaragoza, Spain}
%\email{igor.irastorza@posta.unizar.es}
\date{}

\maketitle

\vspace{-1cm}
\begin{center}
\begin{em}\footnotesize

$^{a}$Laboratory of Nuclear and High Energy Physics, University of
Zaragoza, 50009 Zaragoza, Spain
\\
$^{b}$Dipartimento di Fisica Teorica, Universit\`{a} di Torino and
INFN, Sez. di Torino, Via P. Giuria 1, I-10125 Torino, Italy
\\
\end{em}
\end{center}

\begin{abstract}
The CUORE (\emph{Cryogenic Underground Observatory for Rare Events}) experiment projects to construct and operate an array of
1000 cryogenic thermal detectors of TeO$_{2}$, of a mass of 760 g each, to investigate rare events physics, in particular, double
beta decay and non baryonic particle dark matter. A first step towards CUORE is CUORICINO, an array of 56 of such bolometers,
currently being installed in the Gran Sasso Laboratory. In this paper we report the physics potential of both stages of the
experiment regarding neutrinoless double beta decay of $^{130}$Te, WIMP searches and solar axions.

\vspace{0.5 cm}



\end{abstract}

\section{Introduction: The CUORE project}

Rare event Physics at the low energy frontier is playing a
significant role in Particle Physics and Cosmology. Examples of
such rare phenomena could be the detection of non-baryonic
particle dark matter (axions or WIMPs), supposedly filling a
substantial part of the galactic haloes, or the neutrinoless
double beta decay. These rare signals, if detected, would be
important evidences of a new physics beyond the Standard Model of
Particle Physics, and would have far-reaching consequences in
Cosmology. The experimental achievements accomplished during the
last decade in the field of ultra-low background detectors have
lead to sensitivities capable to search for such rare events. Dark
matter detection experiments have largely benefited from the
techniques developed for double beta decay searches. Both types of
investigation, which have very close relation from an experimental
point of view, are the two main scientific objectives of CUORE.

Due to the low probability of both types of events, the essential
requirement of these experiments is to achieve an extremely low
radioactive background. For that purpose  the use of radiopure
detector components and shieldings, the instrumentation of
mechanisms of background identification, the operation in an
ultra-low background environment, in summary, the use of the
state-of-the-art of low background techniques is mandatory. In
some of these phenomena, like the case of the interaction of a
particle dark matter with ordinary matter, a very small amount of
energy is deposited, and the sensitivity needed to detect events
within such range of energies, relies in how low the energy
threshold of detection is. In addition, to increase the chances of
observing such rare events a large amount of detector mass is in
general advisable, and so most of the experiments devoted to this
type of searches are planning to enlarge their detector mass,
while keeping the other experimental parameters (background,
energy thresholds and resolutions) optimized.

On the other hand, the recent development of cryogenic particle
detection \cite{Mosca} has lead to the extended use of thermal
detectors \cite{Fio84} to take advantage of the low energy
threshold and good energy resolution that are theoretically
expected for the thermal signals. This detection technology has
also the bonus of enlarging the choice of materials which can be
used, either as DM targets or $2\beta$ decay emitters. After a
long period of R\&D to master the techniques used in cryogenic
particle detectors, low temperature devices of various types are
now applied to the detection of double beta decay or particle dark
matter \cite{Mor99}. A major exponent of this development is the
MIBETA (\emph{Milano Double Beta}) experiment \cite{Pirro:2000fi},
which has successfully operated a 20 TeO$_2$ crystal array of
thermal detectors of a total mass of 6.8 kg, the largest cryogenic
mass operated up to now. With the objective of going to larger
detector masses and of improving the sensitivity achieved in the
smaller arrays, the CUORE (\emph{Cryogenic Underground Observatory
for Rare Events}) project \cite{Fiorini:1998gj} was born some
years ago as a substantial extension of MIBETA. The objective of
CUORE is to construct an array of 1000 bolometric detectors with
cubic crystal absorbers of tellurite of 5 cm side and of about 760
g of mass each. The crystals will be arranged in a cubic compact
structure and the experiment will be installed in the Gran Sasso
Underground Laboratory. The material to be employed first is
TeO$_2$, because one of the main goals of CUORE is to investigate
the double beta decay of $^{130}$Te, although other absorbers
could also be used to selectively study several types of rare
event phenomenology. Apart from the MIBETA experiment, already
completed, a wide R\&D program is under way in the framework of
CUORICINO, a smaller and intermediate stage of CUORE, which
consists of 56 of the above crystals (total mass of 42.5 kg).
CUORICINO is currently being mounted in the Gran Sasso Laboratory
(LNGS) and is, by far, the largest cryogenic detector on the
stage. The preliminary tests are
encouraging \cite{Giulani}.% \cite{Pirro:2000fi}.

    In the present work the prospects of CUORE and CUORICINO
experiments with respect to their double beta decay discovery
potential and their detection capability of WIMP and axions are
presented.

\section{Background considerations}

    From the radioactive background point of view, thermal detectors
are very different than, say, conventional ionization detectors,
the former being sensitive over the whole volume, implying that
surface impurities may play an important role. In principle,
bolometric detectors like those of the MIBETA, CUORICINO and CUORE
experiments are expected to have radioactive backgrounds larger
than that of the conventional germanium ionization detectors, also
because the more complex technology of cryodetectors had not been
yet fully optimized from the point of view of the radiopurity of
the components near the detector. On the other hand, the
production of radiopure absorbers for thermal detectors is less
mastered than the production of radiopure Ge crystal. However,
after a considerable R\&D effort a significant improvement in the
radiopurity of bolometers has been accomplished during the last
years, and now these devices have achieved very competitive
backgrounds
\cite{Abusaidi:2000wg,Benoit:2001zu,cresst,Alessandrello2}.

    As one of the main ingredients in the prospective physics
potential of the CUORICINO and CUORE experiments is the level of
background achievable, an evaluation of the expected background
must be done, including the contribution of the possible
background sources. A preliminary estimation was made in the
Status Report of CUORICINO presented to the LNGS Scientific
Committee \cite{Alessandrello}. A more complete estimate is
underway. In the following we will give a briefing of that
(unpublished) reference and will add further considerations.

    There are several background sources to be considered: the
environmental backgrounds of the underground site (neutrons from
the rocks, ambient gamma flux...), the intrinsic radioactivities
of the detectors, components and shielding (bulk and surface), the
cosmic muon induced backgrounds (neutrons, muon direct
interactions \dots), as well as the possible cosmogenic induced
activities produced when the detector and components were outside
the underground laboratory. Even small, the leakage of double beta
counts emitted by the absorber's nuclei in the relevant region of
analysis might be a potential source of background. The depth at
which the experiment will be performed, plus the addition of an
efficient cosmic veto and a suitable passive shielding (lead,
polyethylene, \dots) will effectively reduce the external
background. So we will refer mainly to the intrinsic background,
although we will make some remarks concerning the other sources.
On the other hand, the experience gained in the MIBETA experiment,
where the measured background is used as a test-bench for checking
the MC estimates, has been essential to know which could be the
expectations for CUORICINO. Using the MIBETA results and the
CUORICINO tests a Monte Carlo study of the expected intrinsic
background has been carried out for the CUORICINO (and CUORE)
set-ups. Supposed radioactivity impurities in the bulk of the
tellurite crystals, as well as in the dilution refrigerator and
surrounding shielding were taken into account. Background
suppression due to the anticoincidence between crystals (which
will be significant in the CUORE set up) was worked out. The
effect of the $\alpha$ surface contaminations observed in the
MIBETA crystals was also evaluated and scaled up to CUORICINO. The
simulation was performed with the GEANT code. Alpha, beta and
gamma emissions (supposed to be uniformly distributed and
isotropically emitted) from the natural chains ($^{238}U$,
$^{232}Th$) and from other isotopes ($^{40}K$, $^{60}Co$,
$^{210}Pb$) were included.

In the case of MIBETA, the measurements based on the $\alpha$
peaks of the background have shown that the bulk intrinsic
contamination of U/Th inside the crystals are very low (less than
a few 10$^{-10}$ g/g) and that the main sources of background are
located on the surfaces. According to the method used to clean the
holder of the crystals and to lap the surfaces (and depending on
the radiopurity of the powders used in the lapping), the
background values, in the best of the cases, stand about roughly
half-a-count per keV, kg and year in the 2500-2556 keV (2$\beta$
decay) region, more precisely, b=0.68$\pm$0.06 c/keV/kg/y in the
"old" and b=0.35$\pm$0.14 c/keV/kg/y in the "new" configuration
(labelled respectively after the best lapping operation and a new
mounting system) \cite{Giulani}. A background spectrum of MIBETA
corresponding to the $2\beta$ decay region is shown in Fig. 2 of
Ref. \cite{Alessandrello2}. The addition of a neutron shielding of
10 cm of borated polyethylene does not reduce the background, as
expected if the dominant contribution comes from the surface. With
the above background value the lower limit of the neutrinoless
half-life of $^{130}Te$ obtained is $T_{1/2}^{0\nu}\geq
1.4\times10^{23}$ y, or equivalently, an upper bound for the
Majorana neutrino mass of $\langle m_{\nu} \rangle \lesssim$2 eV
(when using the Heidelberg nuclear matrix element calculation of
Ref. \cite{QRPA1}), which is the second-best published result
\cite{Alessandrello2}. Similarly, as far as the low energy region
is concerned, the most recent current background of MIBETA,
obtained after the new lapping of the crystal detector surfaces,
stands around 1 c/keV/kg/day between 10--50 keV and 0.2
c/keV/kg/day between 50--80 keV, region where the dark matter
signal is expected \cite{Giulani}. Ref. \cite{Giulani} shows the
MIBETA background spectrum corresponding to the low energy, dark
matter region. With these background values, the exclusion plot of
WIMPs interacting coherently with Te and O is depicted as the
(dashed) contour of figure \ref{cuoricino_exclusion}. Both results
of MIBETA, for the low ($<$100 keV) and the high ($\sim$2500 keV)
energy regions will serve as references for the CUORICINO/CUORE
projections.

To reproduce and understand the contribution to the MIBETA
background, a MC calculation has been made for MIBETA (and then
extended to the CUORICINO and CUORE geometries) including the main
components and geometry of the experimental set-up (array of
crystals, holder, cryostat, shielding, \dots). The radioactive
impurities used as inputs were all the contaminants identified in
the MIBETA background spectrum, through the alpha or gamma lines.
They are $^{238}$U in secular equilibrium, $^{214}$Bi, $^{210}$Pb,
$^{232}$Th in secular equilibrium, $^{40}$K and $^{60}$Co. The
amount in which each impurity is present is obtained as the upper
limit for which the MC simulation matches best the measured
background. A previous hint about the amount of such
radioimpurities was obtained by measuring in low background HPGe
detectors (at Gran Sasso and Canfranc underground Laboratories)
the materials and components typically used in the experimental
set-up.

The main results -which should be taken as the expectations for
the CUORICINO background- are the following: in the very low
energy region (from 10 to 100 keV), the background is dominated by
$^{210}$Pb surface contamination located either on the crystal
surface or on the surface of the copper mounting box. This
contamination has been identified through the 46 keV peak and the
5.3 and 5.4 MeV alpha lines. The continuum between 100 and 1500
keV seems to be determined by a $^{40}$K surface contamination of
the crystals/copper mounting box. The continuum counting rate in
the energy region 1500-4000 keV could be explained by a crystal
surface contamination of $^{238}$U, $^{226}$Ra, $^{232}$Th and
$^{210}$Pb. This surface contamination seems to be the dominant
contribution to the background in the 2$\beta$ decay region, and
only a minor contribution should be attributed to the $^{232}$Th
bulk contamination.

The U/Th and $^{210}$Pb surface contaminations of the old MIBETA
crystals (which have been of big concern in the R\&D of CUORICINO)
have been significantly reduced (but not eliminated) by means of a
new lapping procedure of the crystals. These surface
contaminations were estimated by the MC simulation of the MIBETA
set-up and properly checked with the experimental data. They
produced an overall rate of about $\sim$0.1-0.2 c/keV/kg/y in the
2.5 MeV region and 0.2-0.3 c/keV/kg/d in the 10-100 keV energy
region, dominating in both cases over the bulk intrinsic
background. The new cleaning procedure carried out in the $3\times
3 \times 6$ cm$^{3}$ crystals of MIBETA has reduced these values
by more than one order of magnitude.

In fact, in the CUORICINO MC simulation the following inputs have
been assumed: the bulk contamination is the upper limit deduced
for such radioimpurities in the corresponding component of MIBETA.
As far as the surface radioimpurities, they are as follows: the
$^{40}$K is assumed to be the same as in MIBETA. The U/Th surface
contamination has been scaled from that of MIBETA, but taking into
account that they are factors 1.5 and 2 smaller per unit surface
than in MIBETA (as deduced from CUORICINO crystals tests). As far
as the $^{210}$Pb surface contamination in the crystals and/or the
copper, similar values that those derived from MIBETA have been
assumed. Notice however that a reduction of one order of magnitude
(or more) in this background is expected to be obtained and that
very likely possibility will be taken into account to derive the
physics prospects of the experiment.

With the above proviso, the MC simulated contribution to the
CUORICINO intrinsic background (bulk and surface) are
b$_{2\beta}$(bulk)$\sim$2$\times$10$^{-2}$ c/keV/kg/y and
b$_{2\beta}$(surface)$\sim$10$^{-1}$ c/keV/kg/y (from U/Th in the
crystals and from $^{210}$Pb in the crystals and copper) in the
2.5 MeV region and b$_{DM}$(bulk)$\sim$3$\times$10$^{-2}$
c/keV/kg/d and b$_{DM}$(surface)$\sim$5$\times$10$^{-2}$
c/keV/kg/d (from U/Th in crystals) and 2$\times$10$^{-1}$
c/keV/kg/d (from $^{210}$Pb in crystals and copper) in the 10-100
keV region.

Taking into account that the surface contaminations could be
reduced a factor 10, one can expect that the global intrinsic
background of CUORICINO could be in the most simple and
conservative estimate of $\sim$a few $10^{-2}$ c/keV/kg/y around
2.5 MeV and a few $10^{-2}$ c/keV/kg/d in the DM low energy
region.

% The results of the simulation are shown
%in Fig.\ref{simspclow} and \ref{simspchigh} for the low and high
%energy regions respectively.

%Such levels of radioimpurities would have made not advisable the use of such crystals. The R\&D carried out for CUORICINO has
%removed partially this surface problem. In fact, the current $\alpha$ surface contamination of the CUORICINO $5\times 5 \times 5$
%cm$^{3}$ tellurite crystals, properly lapped, stand at a level of $\sim$0.06 (at 3-4~MeV), $\sim$0.12 (at 4-5~MeV) and $\sim$5.4
%(at 5-6~MeV) c/keV/kg/y.

%between 27
%and 40 times lower than before (depending on the energy region).
%stand at a level of which would produce c/keV/kg/d   c/keV/kg/y.

%Concerning the cosmogenic activation, the radionuclides produced
%by the activation of tellurium by cosmic ray neutrons during
%fabrication and transport of the crystals from the factory to the
%underground laboratory are mostly tellurium isotopes, $^{125}$Sb
%and tritium, these last two of some concern because of their long
%half-lives (2.7 and 12.3 years, respectively), but they are still
%below the intrinsic (bulk and surface) background (in particular,
%$\sim$0.1 c/kg/d from $^{3}$H). Other cosmogenically induced
%radionuclides have been disregarded because their lifetimes are
%relatively short.

Concerning the cosmogenic activation of the crystals produced by
cosmic rays when they were above ground (during fabrication and
transportation of the crystals from the factory to the underground
laboratory) we have carried out a Monte Carlo estimation by using
the code COSMOPACK \footnote{We thank Y. Ramachers for providing
us the code used by the CRESST Collaboration}. According with the
history of the MIBETA (and CUORICINO) detectors, we have assumed 2
months of exposure to cosmic rays and then 1 year of cooling down
of the induced activities. The radionuclei produced by the
activation of tellurium are mostly tellurium isotopes
(A=121,123,125,127) as well as $^{124}$Sb, $^{125}$Sb and tritium,
these last two being of more concern because of their half-life
(beta decays of 2.7 years, end-point energy of 767 keV and 12.3
year, end-point energy of 18 keV, respectively). The induced
$^{124}$Sb has also a beta decay (half-life of 60 days, end-point
energy of 2905 keV) and a few gamma lines. The total induced
activities remaining after one year underground have been
estimated to be 6.5 c/kg/d ($^{125}$Sb), 1.8 c/kg/d ($^{124}$Sb)
and 0.1 c/kg/d ($^{3}$H). In the case of tritium, for instance,
the contribution to the event rate (from threshold (5 keV)
onwards) is $5 \times 10^{-3}$ c/keV/kg/d. Their incidence in the
background count rate are
% is about $10^{-2}\sim
%10^{-3}$ c/keV/kg/d, whereas they would produce $\leq 10^{-3}$
%c/keV/kg/y in the region around 2500 keV. They are
still below the intrinsic radioactivities (bulk and surface) of
the crystals but they might set a bound for the achievable
background of the experiment, unless the cooling down period be
extended. No special incidence of the cosmogenic activation has
been seen in the recent MIBETA spectra. However, some lines of the
above mentioned isotopes have been observed when the crystals were
temporarily taken out of LNGS for lapping the surfaces and
operated again.

Notice that, in any case, the cosmogenic activation is already
included in the experimental data of the MIBETA background and so,
it has been taken into account in the extrapolation made in going
from MIBETA to CUORICINO (both types of crystals follow, roughly,
the same history of fabrication in China, transportation and
storage underground).

On the other hand, the leakage of $2\beta$ counts in the low
energy region ($<100$ keV), due to the double beta decay of
$^{130}$Te is totally negligible ($<10^{-4}$ c/kev/kg/d, assuming
for the $^{130}$Te a $2\beta_{2\nu}$ half-life of $\sim 10^{21}$
y, an average of the two current measured results of 7.3$\times
10^{20}$ y and 1.2$\times 10^{21}$ y \cite{Alessandrello2}).

As noticed before, neither cosmic muons nor neutrons have been
taken into account, in detail, in the estimation of the
background. However, the following simplified arguments will serve
to have an approximate idea of which could be their contribution.

The depth of the LNGS (3500 m.w.e) reduces the muon flux down to
$\sim$ 2$\times 10^{-8}$ cm$^{-2}$s$^{-1}$, but a further
effective reduction can be obtained with the use of an efficient
(99.9\%) active veto for not to miss muons traversing the
detectors and to tag possible events associated with them.
Consequently, the muon-induced events contributing to the
background are expected to be only a small component of it. The
main contribution to the expected background, as seen in MIBETA,
is coming from the intrinsic radioimpurities (bulk or surface).

%Notice first that the LNGS site (3500 m.w.e) reduces the muon flux
%down to 2.5$\times 10^{-8}$ cm$^{-2}$s$^{-1}$ from its value $\sim
%10^{-2}$ cm$^{-2}$s$^{-1}$ at sea level.
On the other hand, the shielding will substantially reduce the
event rate due to particles external to the detector from various
sources (neutrons and photons), from radioactivity in the
environment (natural decay chains U/Th, $^{210}$Pb, $^{40}$K,
\dots), as well as muon-induced in the surroundings or in the
shielding itself. The passive shielding typically consists of a
neutron screen (blocks of -borated- polyethylene of 10-20 cm
thickness) to attenuate and moderate neutrons and a shell in lead
(of about 10-15 cm) to attenuate the incoming external photons;
the innermost part is made of archaeological lead (2000 years old)
with a very small content ($<$4 mBq/kg) of $^{210}$Pb (half-life
22 years) and of high radiopurity.
%to avoid
%the bremsstrahlung of $^{210}$Bi (1116 keV end point) and the
%corresponding background photons in the detectors.
The successive shielding barriers, active and passive, guarantee a
very substantial reduction of the external background sources (we
will show how this reduction applies to the CUORICINO case).

Neutrons may constitute a worrisome background in dark matter
experiments because for appropriate neutron energies (few MeV)
they can produce nuclear recoils ($\lesssim $100 keV) in the
detector target nuclei which would mimic WIMP interactions. Simple
kinematics tells that in the case of tellurium, neutrons of 1(5)
MeV could elastically scatter off tellurium nuclei producing
recoils of energies up to 31 (154) keV. In general, one considers
neutrons of two origins: from radioactivity in the surroundings or
muon-induced. Depending on the overburden of the underground site
(i.e., depending on the muon flux), muon-induced neutrons are
produced, at lesser or greater rate, both inside and outside the
shielding. They are moderated (according to their energies) by the
polyethylene/lead shield (when produced outside) or tagged by the
muon veto coincidence (when produced within the passive
shielding). A certain fraction of the neutrons of, say, a few MeVs
produced outside can pass the veto reaching the detector and
producing "dangerous" nuclear recoils and $\gamma$ background.
However, a significant fraction of the associated events can be
vetoed because of their interaction with the veto and also because
of the hadronic showers trained by the muons (see Ref.
\cite{Abusaidi:2000wg}). We have MC-estimated -see later- how many
of these neutrons can punch through the shielding.

In the case of external neutrons (from the rocks, from fission
processes or from (n,$\alpha$) reactions, as well as neutrons
originated by muons in the walls of the underground site), the
ambient neutron flux has been measured in LNGS. The result is of
$\sim1\times 10^{-6}$ cm$^{-2}$s$^{-1}$ for the thermal component,
$\sim2\times 10^{-6}$ cm$^{-2}$s$^{-1}$ for the epithermal and
$\sim2\times 10^{-7}$ cm$^{-2}$s$^{-1}$ for energies over 2.5 MeV
\cite{Belli}. They are fairly well moderated by the polyethylene
and eventually absorbed or captured. We have carried out a Monte
Carlo simulation of the propagation of neutrons through a typical
(but simplified) shielding of CUORICINO. In particular, we have
estimated the distance distributions for neutrons (of energies 1
and 5 MeV) to thermalize (E$<$0.1 eV) in polyethylene. The maxima
of the distributions are at 5 and 7 cm respectively. For instance,
90\% of neutrons of 1 MeV (5 MeV) thermalize after 12 cm (22 cm)
of polyethylene. In a shielding of 40 cm of polyethylene, 99.7 \%
of neutrons of 5 MeV are moderated down to 0.1 eV (as well as
practically all neutrons of 1 MeV). For neutrons of higher
energies the fraction of them which are thermalized after, say, 40
cm of water is $\sim$92\% (for neutrons of 10 MeV), $\sim$83\%
(for neutrons of 25 MeV) and $\sim$50\% (for neutrons of 50 MeV).
Then, taking into account the energies of fission neutrons (rocks,
\dots) and the small number of cosmic muon-induced neutrons at
3500 m.w.e., a neutron shielding of $\sim$ 40 cm of polyethylene
(or water) would practically reduce the external neutron flux down
to energies not dangerous from the point of view of the nuclear
recoils they could produce.

On the other hand, the incoming flux of external neutrons becomes
attenuated in the shielding material (in the polyethylene neutron
moderator or in the gamma lead shielding). As said before, we have
developed a MC simulation of such attenuation as well as the
associated photon production originated by these neutrons (via
inelastic scattering, radiative capture) in the shielding
material. The MC estimate assumes a shielding box of external
dimensions typical of those of CUORE (1 m $\times$ 1 m $\times$ 2
m), with walls of lead and polyethylene of various thickness, to
determine the fraction of incident neutrons (of initial energies
of 1 and 5 MeV) which are able to punch through the thickness of
the shielding box. Taking, for instance, a box consisting of 10 cm
of polyethylene (external) and 10 cm of lead (internal), it turns
out that about 2\%, 4\% and 10 \% of external neutrons having
respectively energies of 0.1, 1 and 5 MeV, are able to punch
through that shielding. The number of neutron-induced photons,
which arrive inside the box (in the detector volume) has been also
evaluated. Starting, for instance, with neutrons of 5 MeV entering
the shielding and reaching the region of the crystals with
energies of 2.6 MeV (generated by inelastic scattering with
$^{208}$Pb) or with 2.2 MeV (due to thermal neutron capture in
hydrogen), they produce $10^{-4}$ photons per neutron (or $8\times
10^{-4}$ photons per neutron) respectively. In the case of
neutrons of 1 MeV, these figures are, respectively, $10^{-4}$
photons per neutron and practically zero photons. These
considerations referred to the typical shielding of the
CUORICINO/CUORE experiment give an idea of the limited incidence
of the external neutrons, as previously stated.
% The fraction of
%neutrons entering the shielding of 10 cm Polyethylene plus 10 cm
%lead with energies of 1 (5) MeV and getting inside are $3 \times
%10^{-2(-1)}$ respectively.

Other neutrons, produced by muon interaction inside the shielding
materials, are very scarce and tagged as events coincident with
the muon veto. As it is well-known, muon-induced neutrons are
originated in a variety of processes. The reduction of muon flux
in underground sites results in a substantial depletion of the
associated neutrons and below $\sim$100 m.w.e. the dominant
sources of neutrons are nuclear fission processes and (n,$\alpha$)
reactions in rocks and other environmental material with sizeable
content of U/Th. The energy spectrum of muon-induced neutrons is
approximated by an inverse energy power law ($E^{-0.88}$ for 1-50
keV and $E^{-1}$ above 50 keV) but other neutron energy spectra
have been proposed.
%referencias del pre-print de FLUKA hep-ex/0101049
The neutron yield per muon can be approximately evaluated through
the simple expression $N_{n}=4.14\times E_{\mu}^{0.74} \times
10^{-6}$ neutrons/(gcm$^{-2}$) per muon \cite{Wang} which fits the
value of 1.5$\times 10^{-4}$ neutrons/(gcm$^{-2}$) of the
muon-induced neutron flux measured by the LVD experiment at Gran
Sasso \cite{Aglietta}. Other authors have used similar
expressions.
% in
%terms of the mean muon energy $\sim$200 GeV, because of the
%$E_{\mu}^{0.74}$ dependence of the neutron yield on the mean muon
%energy.

Using an average neutron yield per muon of 2.2$\times 10^{-4}$
neutrons/(gcm$^{-2}$), together with the LNGS muon flux
(2.5$\times 10^{-8}$ $\mu$/(cm$^{2}$s)), that would produce in the
simplified CUORE shielding referred above (box of $1\times 1\times
2$ m$^{3}$, with walls of polyethylene (10 cm) %, $\sim$670 kg)
and lead (10 cm)), %$\sim$6.3 tons),
about $\sim$0.5 neutrons/day in the polyethylene shield and
$\sim$4 neutrons/day in the lead shell.
%So, roughly, according to the shielding material $2\times 10^{-2}$
%to $2\times 10^{-3}$ neutrons per muon will be produced, or
%equivalently, an average flux of $10^{-10}$ n/cm$^{2}$/s (yet not
%reduced in its way to the array of bolometers) will enter the
%detector area (3$\times$2 m$^{2}$) to produce a negligible number
%of neutron collisions per day (even assuming a few collisions per
%neutron), or equivalently, less than $\sim5 \times 10^{-3}$ c/kg/d
%in the case of CUORE and $\sim2 \times 10^{-2}$ in the case of
%CUORICINO.
So, independently of the mechanism set up to reject or tag the
events associated to neutrons, their rather small number is
expected to play a secondary role in the total background compared
with other, intrinsic, sources of background. On the other hand,
much in the same way, the remarks we made above about the
cosmogenic induced activity when going from MIBETA to CUORICINO
also apply to the neutron induced background, which is already
incorporated to the MIBETA background data, and so taken into
account in the extrapolation. A more complete and detailed study
of the external, non intrinsic background rates is underway.

%An estimate of the attenuation of neutrons produced by a typical
%shielding of CUORICINO made of 10 cm of polyethylene (outside)
%plus 10 cm of lead (inside) has been carried out by using the
%GEANT4 package. Neutrons are supposed to be uniformly distributed
%and isotropically emitted from the external surface of the
%shielding. The outcome of that estimate is that about 2\%, 4\% and
%10 \% of external neutrons having respectively energies of 0.1, 1
%and 5 MeV, are able to punch through that shielding. On the other
%hand, the neutron fluxes measured at Gran Sasso are of
%$\sim1\times 10^{-6}$ cm$^{-2}$s$^{-1}$ for the thermal component,
%$\sim2\times 10^{-6}$ cm$^{-2}$s$^{-1}$ for the epithermal and
%$\sim2\times 10^{-7}$ cm$^{-2}$s$^{-1}$ for energies over 2.5 MeV
%\cite{Belli}.

The $\gamma$-ambient background in Gran Sasso is $\sim$ 1
cm$^{-2}$s$^{-1}$ \cite{Arpesella} and it can be attenuated by a
proper lead shielding.

The background values quoted for MIBETA were obtained nevertheless
with a more simplified shielding (no active veto, no neutron
shielding except in the last running, where 10 cm of polyethylene
was added). In spite of this fact, the MIBETA background
\cite{Giulani} at low energy ($\sim$1 c/keV/kg/d at threshold -10
keV- and at 3500 m.w.e.) is similar than the measured event rate
anticoincident with the veto in the CDMS experiment
\cite{Abusaidi:2000wg} ($\sim$ 2 c/keV/kg/d at 10 keV and at 20
m.w.e.) and in the EDELWEISS experiment
\cite{Benoit:2001zu}($\sim$ 1.8 c/keV/kg/d at 30 keV and 4000
m.w.e.), without veto, (in both cases, obviously, prior to
charge-heat discrimination) and is roughly equal to that of DAMA
\cite{Bernabei} and ANAIS \cite{anais} ($\sim$ 1.5 c/keV/kg/d, at
2 keV and 3500 m.w.e. and at 4 keV and 2450 m.w.e. respectively)
but still one order of magnitude worse than that of IGEX
\cite{IGEXDM} ($\sim$ 0.2-0.05 c/keV/kg/d at 4 keV and 10 keV
respectively, and at 2450 m.w.e.). In the double beta decay region
($\sim$ 2.5 MeV) the MIBETA background values ($\sim$ 0.3-0.6
c/keV/kg/y in, respectively, the old and new set-up) are
competitive but still higher than those of the IGEX experiment
\cite{IGEX2beta} (0.05 c/keV/kg/y at 2 MeV) (which uses Pulse
Shape Discrimination).

The challenge of CUORICINO (and later on, of CUORE) is to reduce
significantly the MIBETA background values (both in the low and
high energy regions) by, say, two orders of magnitude without
using background discrimination mechanisms like the simultaneous
measurement of charge (or light) and heat. We argue that it can be
achieved in two steps.

Starting from the current background values of MIBETA ($\sim$ 1
c/keV/kg/d from 10-50 keV, $\sim$ 0.2 c/keV/kg/d between 50-80 keV
and $\sim$ 0.3 c/keV/kg/y around 2.5 MeV) and taking into account
the preceding discussion about the incidence of the various
components of the background, we expect that these background
values can be reduced by one to two orders of magnitude, going
down to values smaller than 0.1-0.05 c/keV/kg/day in the low
energy region and to about 0.01 c/keV/kg/y (or lower) in the
double beta region, as suggested by the MC simulation and the
nature and location of the background radioactivities. That
extrapolation is very conservative if one takes into account the
exhaustive radiopurity selection of the detector materials and
components which is being done for CUORICINO, the new lapping
procedure of the crystal surface, the closer packing of the
crystals, the suppression of significant amount of holding
material, the powerful anticoincidence rejection provided by the
segmented geometry of the crystal array and the use of a much
better shielding than that employed in MIBETA. Nevertheless, as a
first step in the background achievements, we will assume in the
following prospective analysis of the CUORICINO capabilities, even
more conservative background values (including intrinsic and
external), more precisely of 1 c/keV/kg/day and of 0.1
c/keV/kg/day in the low energy region (at threshold, $\sim$5 keV)
and values of 0.1 c/keV/kg/y and of 0.01 c/keV/kg/y in the 2--2.5
MeV region. The forthcoming results of the CUORICINO experiment
will tell us how much this reduction can be further pursued. Even
at these conservative values, CUORICINO is, as we will proof in
the following, a powerful instrument to look for $2\beta$ decays
and WIMPs. In the case of CUORE the background values that will be
assumed (also approximatively) are 0.1 and 0.01 c/keV/kg/day in
the low energy region and 0.1 and 0.01 c/keV/kg/y around 2.5 MeV,
which should be considered, also, two successive steps of the
experiment.

Regarding the expected threshold and resolution, in the CUORICINO
tests \cite{Pirro:2000fi} energy thresholds of $\sim 5$ keV have
been obtained and a resolution of 1 keV at the 46 keV line of
$^{210}$Pb achieved, in some of the detectors. Ref. \cite{Giulani}
illustrates this point. In the case of CUORE thresholds of 5 keV
and energy resolutions of 1 keV at low energies will be assumed.
As far as the energy resolutions obtained in the double beta decay
region, values of 3 keV at 2615 keV were achieved in some crystals
but they are worse (by a factor two) in others. See Fig. 1 of Ref.
\cite{Alessandrello}. Nevertheless, they are clearly better than
that obtained in the previous $6\times 3 \times 3$ cm$^{3}$
(MIBETA) crystals (8$\sim$10 keV). Taking into account these
expectations, we discuss in the following the prospects of
CUORICINO and CUORE for double beta decay searches (section 2),
for WIMP detection (section 3) and for solar axion exploration
(section 4).

\section{Double beta decay}

One of the main scientific objectives of the CUORE detectors is to
search for the double beta decay of the $^{130}$Te isotope
contained in the (natural) TeO$_2$ set of crystals.

The importance of the nuclear double beta decay as an invaluable
tool to explore particle physics beyond the Standard Model has
been repeatedly emphasized and widely reported \cite{doblebeta}.
In the Standard Model of Particle Physics neutrinos are strictly
massless, although there is no theoretical reason for such a
prejudice. On the experimental side, moreover, there exist strong
evidences from atmospheric neutrino data (from SuperKamiokande)
and from experiments with solar neutrinos from Homestake, Gran
Sasso and Kamioka, since long time ago, which suggest that
neutrinos have indeed masses and oscillate among the three
species. The recent results of the solar $\nu$ experiment from SNO
with both CC (Charged Current) and NC (Neutral Current)
interactions \cite{Ahmad:2001an}, also combined with SuperK have
provided, definitively, a strong evidence that neutrinos do
oscillate and, consequently, the existence of non-zero mass
neutrinos. However, neutrino oscillation experiments provide the
squared mass difference between the neutrino species but not their
absolute value and scale. The neutrinoless double beta decay would
help to solve this question and to disentangle the hierarchy
scheme of the neutrino flavours. Most of the models (see
\cite{pascoli}) indicate that the Majorana neutrino mass parameter
could be around (or slightly below) $\langle m_{\nu}\rangle \sim
0.05$ eV, value within reach of the future double beta decay
experiments, like CUORE. On the other hand, galaxy formation
requires a small amount of hot non-baryonic dark matter likely in
form of neutrinos to match properly the observed spectral power at
all scales of the universe. The question of the neutrino mass is
one of the main issues in Particle Physics.

In the Standard Model, neutrinos and antineutrinos are supposed to
be different particles, but no experimental proof has been
provided so far. The nuclear double beta decay addresses both
questions: whether the neutrinos are self-conjugated and whether
they have non-zero Majorana masses. In fact, the lepton number
violating neutrinoless double beta decay
$(A,Z)\rightarrow(A,Z+2)+2e^-$ ($2\beta 0\nu$) is the most direct
way to determine if neutrinos are Majorana particles. Moreover,
the observation of a $2\beta 0\nu$ decay would imply a lower bound
for the neutrino mass, i. e. at least one neutrino eigenstate has
a non-zero mass.

Another form of neutrinoless decay, $(A,Z)\rightarrow(A,Z+2)+2e^-
+ \chi$ may reveal also the existence of the Majoron ($\chi$), the
Goldstone boson emerging from the spontaneous symmetry breaking of
$B-L$, of most relevance in the generation of Majorana neutrino
masses and of far-reaching implications in Astrophysics and
Cosmology. These and other issues, make the search for the
neutrinoless double beta decay an invaluable tool of exploration
of non-standard model physics, probing mass scales well above
those reached with accelerators. That is the motivation why there
are underway dozens of experiments looking for the double beta
decay of various nuclei \cite{doblebeta} like $^{76}Ge$ (IGEX
\cite{IGEX2beta}, Heidelberg-Moscow \cite{HM2beta}), $^{100}Mo$
and others (NEMO \cite{sarazin}, ELEGANTS \cite{elegants}) and
$^{130}Te$ (MIBETA, CUORICINO) and a few big experimental
projects, like CUORE, Majorana \cite{majorana} ($^{76}Ge$), MOON
\cite{ejiri} ($^{100}Mo$) and EXO \cite{danilov} ($136Xe$).

The cryogenic thermal detectors provide new double beta emitter
nuclei to be explored in "active" source=detector calorimeters.
Some of them have been tested and others are already in running
detectors, like $^{48}$Ca in CaF$_2$, $^{130}$Te in TeO$_2$, and
$^{116}$Cd in CdWO$_4$. As far as the Tellurium Oxide is
concerned, the 130-Tellurium isotope is a good candidate for
double beta decay searches: its isotopic content in natural
Tellurium is 33.87\%, and its 2$\beta$ Q-value
($Q_{2\beta}=2528\pm1.3$ keV) is reasonably high to escape from
the main radioimpurity lines when looking for a neutrinoless
signal. Moreover, this Q-value happens to be between the peak and
the Compton edge of the 2615 keV line of $^{208}$Tl, which leaves
a clean window to look for the signal. Finally, it has a fairly
good neutrinoless nuclear factor-of-merit
$F_N^{0\nu}=G_{0\nu}\left|M^{0\nu}\right|^2$, where $G_{0\nu}$ is
an integrated kinematical factor qualifying the goodness of the
$Q_{2\beta}$ value (large phase space) and $M^{0\nu}$ the
neutrinoless nuclear matrix elements characterizing the likeliness
of the transition.

In table \ref{tab_doblebeta} we quote the values of $F_N^{0\nu}$
in the $^{130}$Te case calculated in various nuclear models
\cite{doblebeta}, together with those of other emitters used in
source=detector calorimeters. It can be seen that no matter the
nuclear model used to compute the neutrinoless decay matrix
elements, the merits of $^{130}$Te are a factor 5--10 more
favorable than those of $^{76}$Ge (the emitter where the best
neutrinoless double beta decay half-life limits have been achieved
so far), which translates into a factor 2 to 3 better as far as
the $\langle m_{\nu} \rangle$ (Majorana neutrino mass parameter)
bounds are concerned.

The detector factor-of-merit $F_D^{0\nu}$, or detection
sensitivity, introduced long time ago by Fiorini, provides an
approximate estimate of the neutrinoless half-life limit
achievable with a given detector. For source=detector devices, it
reads:

\begin{equation}\label{merit}
F_D^{0\nu} = 4.17\times 10^{26} \times \frac{a}{A}
\sqrt{\frac{Mt}{b\Gamma}}\times \epsilon {\rm \ \ \ years}
\end{equation}

\noindent where $A$ is the atomic mass, $a$ is the isotopic
abundance, $M$ the detector mass in kg, $b$ the background in
c/keV/kg/y in the 2$\beta$ neutrinoless decay region, $t$ the
running time in years, $\Gamma$ the FWHM energy resolution in keV
and $\epsilon$ the detector efficiency (which is practically one
within the fiducial volume of the detector). In the case of a
TeO$_2$ crystal detector, $F_D^{0\nu} \sim 8.86 \times 10^{23}
\sqrt{\frac{Mt}{b\Gamma}}$, with $M$ the crystal mass in kg and
$b$ the background in counts per keV and year per kg of detector
mass.

The simplest projection for CUORICINO uses a background of b=0.3
c/keV/kg/y (at 2.5 MeV) and an energy resolution FWHM(2.5
MeV)$\sim$ 8 keV, which are the performances obtained currently in
MIBETA (notice that the background in the "old" MIBETA version was
of $\sim$0.6 c/keV/kg/y). With that proviso, the sensitivity of 1
year exposure of CUORICINO (42.5 kg mass of crystals) will be
$T_{1/2}^{0\nu}\geq 3.7\times10^{24}$ y. As far as the $\langle
m_{\nu}\rangle$ limit is concerned (given as usually by $\langle
m_\nu \rangle \leq m_e / \sqrt{F_D^{0\nu}F_N^{0\nu}}$ ), since the
nuclear factor-of-merit of $^{130}$Te is larger than that of
$^{76}$Ge (see Table \ref{tab_doblebeta}) the $^{130}$Te emitter
provides a factor 2--3 better $\langle m_{\nu} \rangle$ bound than
that obtained from $^{76}$Ge for the same half-life limit. In
other words, one needs to reach a half-life limit of only
$T_{1/2}^{0\nu}>5\times10^{24}$ years to get the same $\langle
m_{\nu} \rangle$ upper bound currently achieved in the best
Germanium experiments ($1.6-1.9 \times10^{25}$ years respectively
\cite{IGEX2beta} and \cite{HM2beta}). In terms of $\langle m_{\nu}
\rangle$ bound, the above simplest projection of
$T_{1/2}^{0\nu}\geq 3.7\times10^{24}$ y means $\langle m_{\nu}
\rangle < 0.36-0.21$ eV using \cite{QRPA1} or \cite{WCSM}, i.e.,
values comparable to, or better than, the best currently achieved
with Ge experiments.

    A second, still very conservative projection would be provided
in the case of reaching a background of b=0.1 c/keV/kg/y (as
discussed in Section 2) and a resolution of $\Gamma=$ 5 keV in the
2.5 MeV region.
%reasonably expected (by
%extrapolating the MIBETA results --new crystal etching-- and
%supported by the MC estimates which include the crystal surface
%contamination), due also to the better quality of the CUORICINO
%crystals,
Such performances can be easily achieved in the very first stage
of CUORICINO (notice that MIBETA has already achieved 0.3
c/keV/kg/y and that the energy resolution in the tests of the
CUORICINO crystals ranges from $\Gamma=3$ keV to $\sim6$ keV at
2615 keV, i.e., notably better than that of smaller $3\times
3\times 6$ cm$^{3}$ of the 20 crystal array of MIBETA). In that
case, one would have $F_D\sim 1.25\times 10^{24} \sqrt{Mt}$ years,
with $M$ (kg) the mass of the Tellurite crystal array. For the
mass of CUORICINO (M=42.65 kg), one will have a sensitivity of
$F_D\sim 8.15\times10^{24} \sqrt{t}$ years. Using a typical
average value of $F_{N}=4\times 10^{-13}$ y$^{-1}$, as obtained in
QRPA \cite{QRPA1,QRPA2,QRPA3,QRPA4,QRPA5}, CUORICINO will have a
mass bound sensitivity of $\langle m_{\nu} \rangle <0.28$ eV in
one year, in the least favourable case. Using, in particular, (for
comparison purposes) the value of $F_{N}=5.33\times 10^{-13}$
y$^{-1}$ of Ref. \cite{QRPA1} which is usually employed in the
$\langle m_{\nu} \rangle$ bound ($\sim$0.33 eV) derived from the
Ge experiments, CUORICINO would provide with these assumptions
$\langle m_{\nu} \rangle <$0.24 eV.

    To go further, one needs to increase the mass of TeO$_{2}$
(CUORE 760 kg) and to reach even a lower background, which the
anticoincidence capability of the CUORE array of one thousand
crystals and the surface radiopurity could, likely, achieve going
down to, say, b=0.05-0.01 c/keV/kg/y.
%, i.e., a factor 6 to 30
%improvement from the latest results of MIBETA (b=0.3 c/keV/kg/y in
%the "new" set-up), or a factor 2 to 10 improvement of the
%background which supposedly will be obtained in the second step of
%CUORICINO (assumed to be b=0.1 c/keV/kg/y in the first step).
In the case of the full CUORE detector, formed by one thousand
crystals of a total mass of 760 kg, even keeping, conservatively,
$b$ and $\Gamma$ as above (b=0.1 c/keV/kg/y, $\Gamma$(2.5 MeV)=5
keV), one would get $T_{1/2}^{0\nu}\geq 3.4\times10^{25} \sqrt{t}$
years, which in one year of statistics would provide $\langle
m_{\nu} \rangle$ bounds ranging from 0.12 eV \cite{QRPA1}, 0.14 eV
\cite{QRPA3}, 0.26 eV \cite{QRPA5} or 0.07 eV \cite{WCSM} just to
mention a few nuclear matrix element estimates. However, the R\&D
to be carried out in CUORE and the operation in the
antincoincidence mode is expected to provide better
figure-of-merit than the values used in the previous predictions.
According to such expectatives, values of $b\sim 0.01$ c/keV/kg/y
and a FWHM energy resolution of $\Gamma \sim 1$ keV will be used
for the last step of CUORE, i.e., a detection sensitivity of
$F_D\sim 8.86\times10^{24} \sqrt{Mt}$ years, from where one can
work out the best expectatives of the whole CUORE ($M=760$ kg of
tellurite), i.e., $F_D\sim 2.5\times10^{26} \sqrt{t}$ years, or
$\langle m_{\nu} \rangle$ bounds ranging from $\sim 0.05~t^{-1/4}$
eV (in \cite{QRPA1,QRPA2,QRPA3,QRPA4,QRPA5,OEM}) to $\sim
0.03~t^{-1/4}$ eV (in \cite{WCSM,GenSen}). The ultimate
sensitivity of CUORE (as far as the Majorana neutrino mass bound
is concerned) stagnates at $\sim 0.05$ eV with a very softened
dependence with time.

\section{WIMP detection}

Recent cosmological observations \cite{Bachcall:1999} provide
compelling evidence for the existence of an important component of
non-baryonic cold dark matter in the Universe. Among the
candidates to compose this matter, Weakly Interacting Massive
Particles (WIMPs) and axions are the front runners. The lightest
stable particles of supersymmetric theories, like the neutralino
\cite{Jungman:1996df}, describe a particular class of WIMPs.

Under the hypothesis of WIMPS as main component of the dark
matter, these particles should fill the galactic haloes and
explain the flat rotation curves which are usually observed in
many galaxies. The detection of such particles could be attempted
both by means of direct and indirect methods. The direct detection
of WIMPs relies on the measurement of their elastic scattering off
the target nuclei of a suitable detector\cite{Mor99}. The non
relativistic and heavy (GeV -- TeV) WIMPs could hit a detector
nucleus producing a nuclear recoil of a few keV. Because of the
small WIMP-matter interaction cross sections the rate is extremely
low. In the case of SUSY WIMPs most of the cross section
predictions \cite{Ellis,Bottino:2001jx,Bergstrom} (derived using
MSSM as the basic frame implemented with different unification
hypothesis) encompass a range of values several orders of
magnitude wide (the so-called scatter plots) providing rates
ranging from 1 c/kg/day down to 10$^{-5}$ c/kg/day according to
the particular SUSY model.

It is well known that the predicted signal for the WIMP elastic
scattering has an exponentially decaying energy dependence, hardly
distinguishable from the background recorded in the detector. The
simple comparison of the theoretical WIMP spectrum with the one
experimentally obtained, provides an exclusion (at a given
confidence level), as dark matter component of the halo, of those
WIMPs with masses ($m$) and cross sections on nucleons ($\sigma$)
which yield spectra above the measured experimental rate. To claim
a positive identification of the WIMP, however, a distinctive
signature is needed. The only identification signals of the WIMP
explored up to now are provided by the features of the Earth's
motion with respect to the dark matter halo. In particular, the
annual modulation \cite{Drukier:1986tm} is originated by the
combination of the motion of the solar system in the galactic rest
frame and the rotation of the Earth around the Sun. Due to this
effect, the incoming WIMP's velocities in the detector rest frame
change continuously during the year, having a maximum in summer
and a minimum in winter. Therefore the total WIMP rate changes in
time with an oscillating frequency which corresponds to an annual
period and a maximum around the beginning of June.

The relative variation of the signal is small (a few percent) so
in order to detect it one needs large detector masses to increase
statistics and several periods of exposure to minimize
systematics. Several experiments have already searched for this
effect \cite{modza,modarg,damaxe} and since 1997 one group has
reported a positive signal \cite{Bernabei} which has been
appearing along four yearly periods. The present situation is no
doubt exciting: on one hand that result has triggered an intense
activity in the field; on the other, the experimental
sensitivities of various types of underground detectors are
entering the supersymmetric parameter space \cite{Bottino:2001jx}
and in particular three of them
\cite{Abusaidi:2000wg,Benoit:2001zu,IGEXDM} are excluding, to a
larger or shorter extent, the region of mass and cross-section
where the reported WIMP is supposed to exist. New data from one of
them \cite{Benoit:2001zu} have excluded totally the DAMA region.
We will discuss in the following the capabilities of CUORICINO and
CUORE to exclude WIMPs using the total time-integrated
experimental rate and comparing it with the predicted nuclear
recoil rate. To look for the annual modulation signal in CUORICINO
/ CUORE experiments, which in principle have enough masses to be
sensitive to it, one needs to know their stability performances.
The analysis of the CUORE / CUORICINO potential for annual
modulation searches will be performed -following statistical
consideration- (see Ref. \cite{Cebrian:2001qk}), with the proviso
that systematic uncertainties are under control. Data on the
stability of MIBETA and on the first running of CUORICINO will be
crucial to asses such hypothesis.

To calculate the theoretical WIMP rate, standard hypothesis and
astrophysical parameters are assumed, i.e., that the WIMPs form an
isotropic, isothermal, non-rotating halo (the isothermal sphere
model) of density $\rho=0.3 $ GeV/cm$^3$, which has a maxwellian
velocity distribution with $v_{rms}=270$ km/s (with an upper cut
corresponding to an escape velocity of 650 km/s), and a relative
Earth-halo velocity of $v_r=230$ km/s). Other, more elaborated
halo models, which have been considered recently \cite{halomodels}
would lead to different results. The same applies when other
astrophysical parameters are employed or when uncertainties in the
halo WIMPs velocity distribution are included \cite{astropar}. The
theoretical predicted rate is expressed in terms of the mass and
cross-section of the WIMP-matter interaction. The cross sections
are normalized per nucleon assuming a dominant scalar interaction,
as is expected, for instance, for one of the most popular dark
matter candidates, the neutralino:

\begin{equation}\label{norm}
    \sigma_{N\chi} = \sigma_{n\chi}A^2 \frac{\mu^2_{W,N}}{\mu^2_{W,n}}
\end{equation}

\noindent where $A$ is the target (oxigen and tellurium) mass number, $\mu^2_{W,N}$ is the WIMP-nucleus reduced mass, and
$\mu^2_{W,n}$ the WIMP-nucleon reduced mass. The Helm parameterization \cite{Helm} is used for the scalar nucleon form factor.
The $(m,\sigma)$ exclusion plot is then derived by requiring the theoretically predicted signal for each $m$ and $\sigma$ in each
energy bin to be less than or equal to the (90\% C.L.) upper limit of the (Poisson) recorded counts. The bin width is assumed to
be equal to the detector resolution.
% Notice that other statistical criteria may lead to somehow different exclusions.

\begin{figure}[h]
\begin{center}
\hspace{0 cm} \psfig{figure=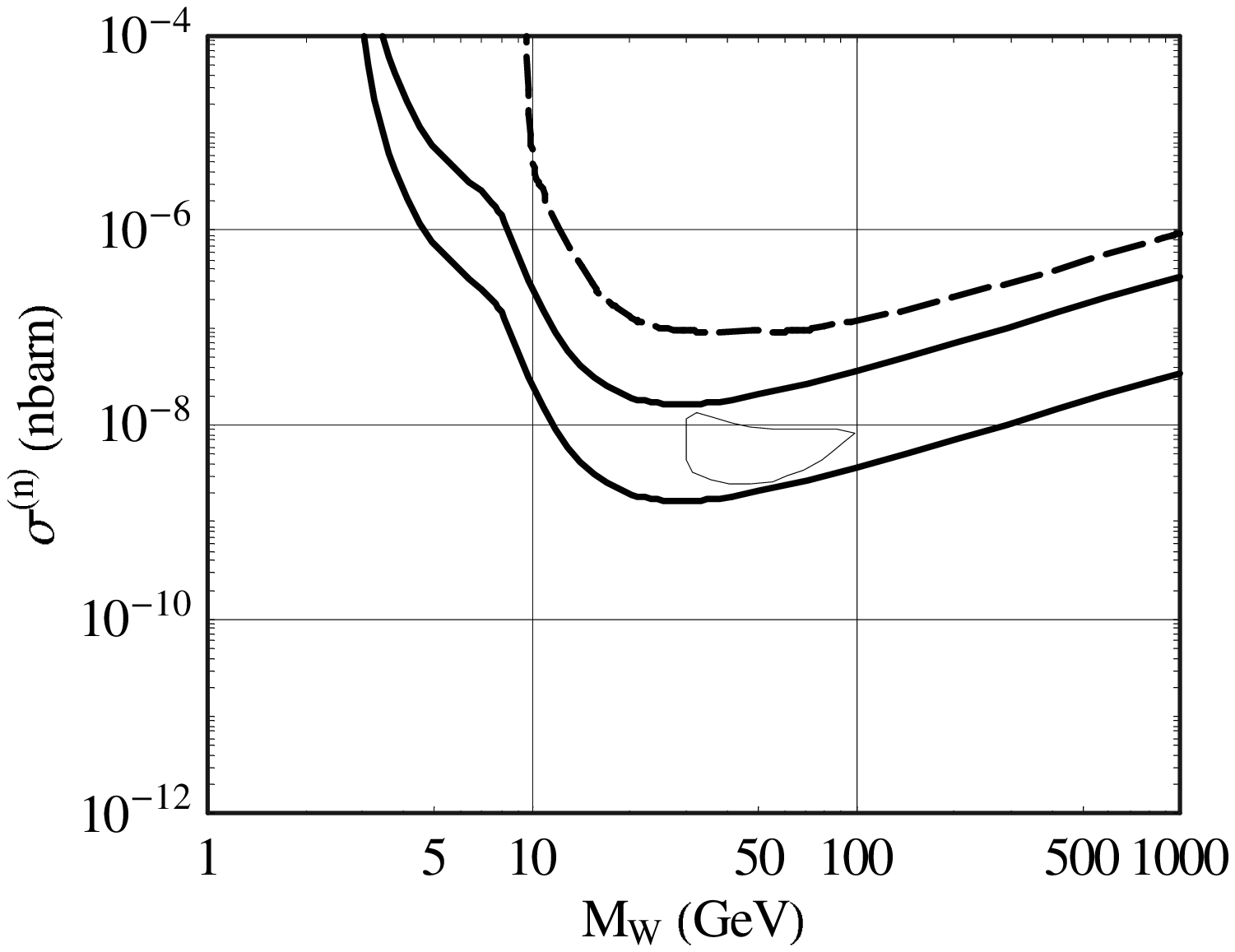,width=120mm}
\caption{Exclusion projected for 2 years of CUORICINO assuming a
threshold of 5 keV, a low energy resolution of 1 keV, and low
energy background levels of 1 and 0.1 c/keV/kg/day respectively.
The closed curve represents the DAMA region. The dashed line
corresponds to the current MIBETA result.
\label{cuoricino_exclusion}}
\end{center}
\end{figure}

\begin{figure}[h]
\begin{center}
\hspace{0 cm} \psfig{figure=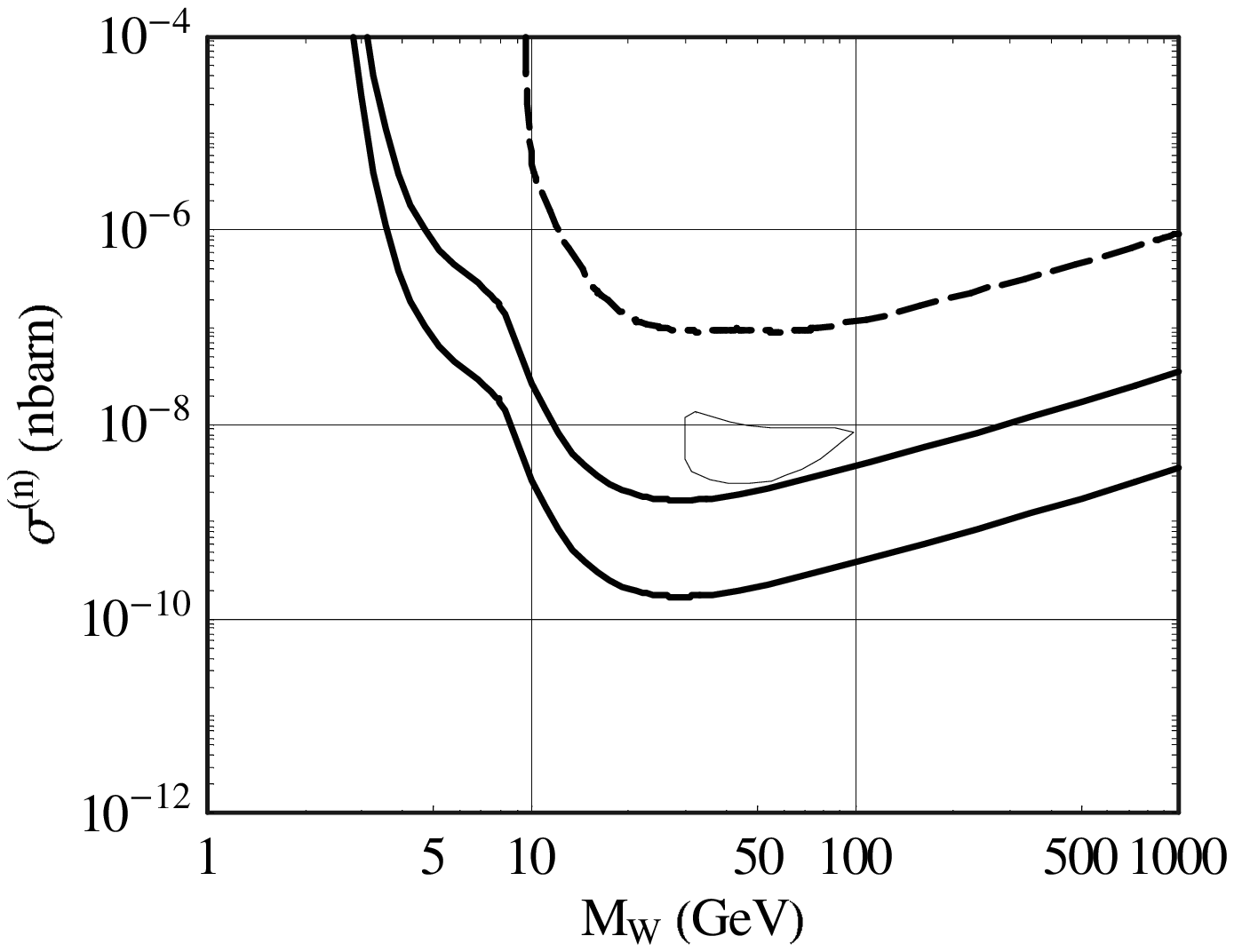,width=140mm}
 \caption{Exclusion projected for 1 year of CUORE
assuming a threshold of 5 keV, a low energy resolution of 1 keV, and low energy background levels of 0.1 and 0.01 c/keV/kg/day
respectively. The closed curve represents the DAMA region. The dashed line corresponds to the current MIBETA result.
\label{cuore_exclusion}}
\end{center}
\end{figure}

In figure \ref{cuoricino_exclusion} the exclusion plots for coherent spin-independent WIMP-matter interaction are shown for two
possible values of the background of CUORICINO, 1 and 0.1 c/keV/kg/day. The first value is of the order of the background already
achieved from threshold onwards (10-50 keV) in the MIBETA latest results (see \cite{Giulani}). The value 0.1 c/keV/kg/day is a
one-order-of-magnitude extrapolation from that currently achieved in MIBETA (see discussion on Section 1) and is close to the one
obtained above 50 keV. The main challenge of this hypothesis is to get b=0.1 c/keV/kg/day below 10 keV. Notice that these values
are more conservative than that derived from the MC simulation, and could be taken as moderate extrapolations of the background
of MIBETA and of the CUORICINO test crystals. In the case of CUORE, background values of 0.1 and 0.01 c/keV/kg/day will be
assumed. Notice, moreover, that values of a few 0.01 have been obtained above 10 keV in the raw spectra of Germanium experiments
(like IGEX \cite{IGEXDM}) without using mechanisms of background rejection, and so it does not seem impossible to achieve such
equivalent ($\leq 0.04$ c/keV/kg/day) small values in crystal thermal detectors of Tellurium (only phonons, and no discrimination
mechanism). To draw the two exclusion contours of Fig. \ref{cuoricino_exclusion}, a low energy resolution of 1 keV and an energy
threshold of 5 keV have been assumed as well as an exposure of 2 years of CUORICINO (84 kg$\cdot$year). The projected exclusion
contours are compared with the one currently obtained from MIBETA (dashed line). In figure \ref{cuore_exclusion}, the exclusions
for the two quoted values of the background of CUORE, 0.1 and 0.01 c/keV/kg/day, are similarly presented for an exposure of 1
year (760 kg$\cdot$year). In both Figures \ref{cuoricino_exclusion} and \ref{cuore_exclusion}, the region corresponding to the
annual modulation positive signal reported by the DAMA collaboration is depicted as the closed "triangular" contour. One can see
that the DAMA region could already be explored by the CUORICINO experiment with a background of 0.1 c/keV/kg/day. Of course CUORE
will be able to explore a larger region of the ($m,\sigma$) plane entering substantially into the scatter plots of the
theoretical predictions of the various SUSY models (see \cite{Bottino:2001jx} and references therein). The reader could make his
own choice according to the credibility he might give to the above extrapolation of the background and other parameters from the
current obtained performances.

As previously noted, CUORE and to some extent CUORICINO have
detector masses large enough to search for the annual modulation
signal. As it is well known, an essential requirement to estimate
the prospects of any detector to search for annual modulation is
to have a superbe control of systematic errors and to assure that
the stability of the various experimental parameters, which might
mimic periodic variations of the signals, are kept within a small
percent of the (already tiny) expected signal. The various changes
of set-up, crystals and shielding of the MIBETA experiment have
not provided a definitive estimation of the long-term stability
parameters of MIBETA. Possible instabilities are that of the
electronic gain and the ensuing time fluctuation of the energy
scale (both in energy thresholds and energy resolutions), the
temperature variations, the possible fluctuation in time of the
efficiency with which the triggered noise is rejected and others.
They must be kept well below the small expected seasonal
modulation of the WIMP signal. The fact that we are dealing with a
very small signal depending on time, which typically amounts to a
fraction between 1$\%$ and 7$\%$ of the average count rates,
reinforces the need for a control of the stability of the
experiment (well below that range) over long periods of time.
%Such superbe stability must be required to any experiment which claims
%to be sensitive to the annual modulation signal, and so should be
%done for CUORICINO and CUORE.
In the case of assuming that all
these fluctuations are controlled well below the levels needed
($<$1\%), then one can proceed to analyze the sensitivity of
CUORICINO/CUORE to the annual modulation signal on purely
statistical grounds. This has been first attempted in
\cite{Ramachers} and \cite{Hasenbalg}, but a more extensive and
rigorous approach is followed in ref. \cite{Cebrian:2001qk} where
sensitivity plots for several types of detectors (and experimental
parameters) are presented, and in particular, for CUORE and
CUORICINO.

    The sensitivity of a given experimental device to the annual modulation signal
(according to the detector material employed and the experimental
parameters of the detectors) has been extensively studied in Ref.
\cite{Cebrian:2001qk} on purely statistical grounds.
%This has been
%first attempted in \cite{Ramachers} and \cite{Hasenbalg}, but a
%more extensive and rigorous approach is followed in ref.
%\cite{Cebrian:2001qk} where sensitivity plots for several types of
%detectors (and experimental parameters) are presented.
Following the guidelines of that reference, it can be precisely
quantified by means of the $\delta$ parameter, defined from the
likelihood function or, equivalently, from the $\chi^2$ function
of the cosine projections of the data (for further details see
ref. \cite{Cebrian:2001qk}):

\begin{equation}\label{eq:asimptotic}
  \delta^2=y(\sigma=0)-y_{min}\simeq\chi^2(\sigma=0)-\chi^2_{min}.
\end{equation}

This parameter measures the statistical significance of the
modulation signal detected in an experimental set of data.
However, for a given ($m,\sigma$) and a given experiment the
expected value $\langle \delta^2 \rangle$ can be estimated using
the expression derived in ref. \cite{Cebrian:2001qk}:

\begin{equation}\label{expansion1}
  \langle \delta^2 \rangle=\frac{1}{2}\sum_{k}
  \frac{S_{m,k}(\sigma,m_W)^2\Delta
  E_k}{b_{k}+S_{0,k}}MT\alpha+2\label{eq:magic}.
\end{equation}

\noindent where $S_{m,k}$ and $S_{0,k}$ are the modulated and
non-modulated parts of the WIMP signal in the $k$th energy bin of
$\Delta E_k$ width, $b_k$ is the background in that energy bin and
$MT\alpha$ the effective exposure, being $\alpha$ a coefficient
accounting for the temporal distribution of the exposure time
around modulation maxima and minima ($\alpha = 1/n \sum_{i=1}^n
\cos^2 \omega (t_i -t_0)$ for $n$ temporal bins).

\begin{figure}[h]
\begin{center}
\hspace{0 cm} \psfig{figure=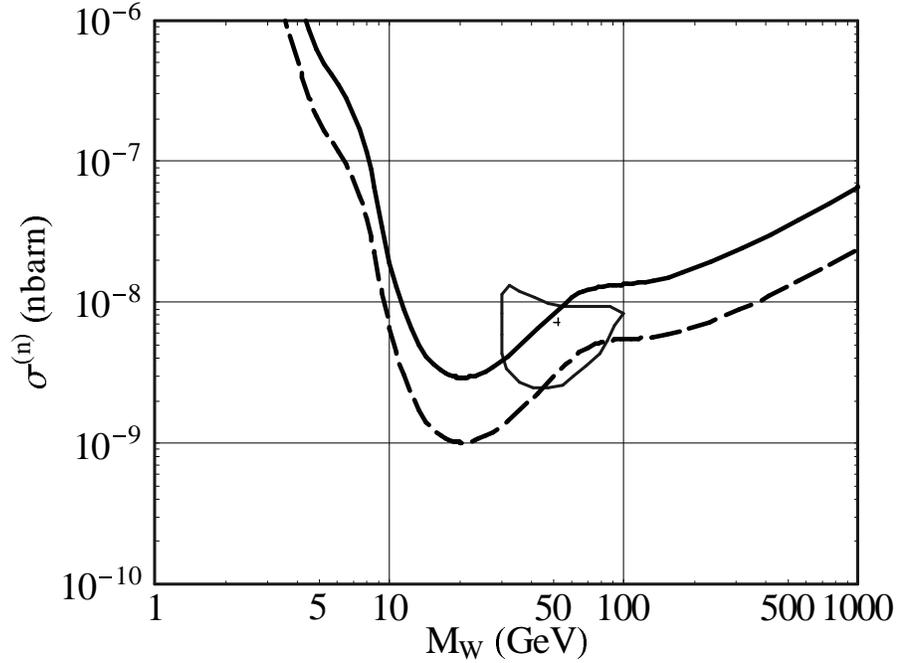,width=120mm}
\caption{Sensitivity plot in the ($m,\sigma$) plane for CUORICINO,
assuming a threshold of 5 keV, flat background $b=$ 1 (solid line)
and 0.1 c/keV/kg/day (dashed line) and two years of exposure (84
kg year). It has been calculated for $\langle\delta^2\rangle=$ 5.6
(see the text). The closed contour represents the 3$\sigma$ CL
region singled out by the modulation analysis performed by the
DAMA experiment \cite{Bernabei}, and the cross indicates the
minimum of the likelihood found by the same authors.
\label{cuoricino_mod}}
\end{center}
\end{figure}

\begin{figure}[h]
\begin{center}
\hspace{0 cm} \psfig{figure=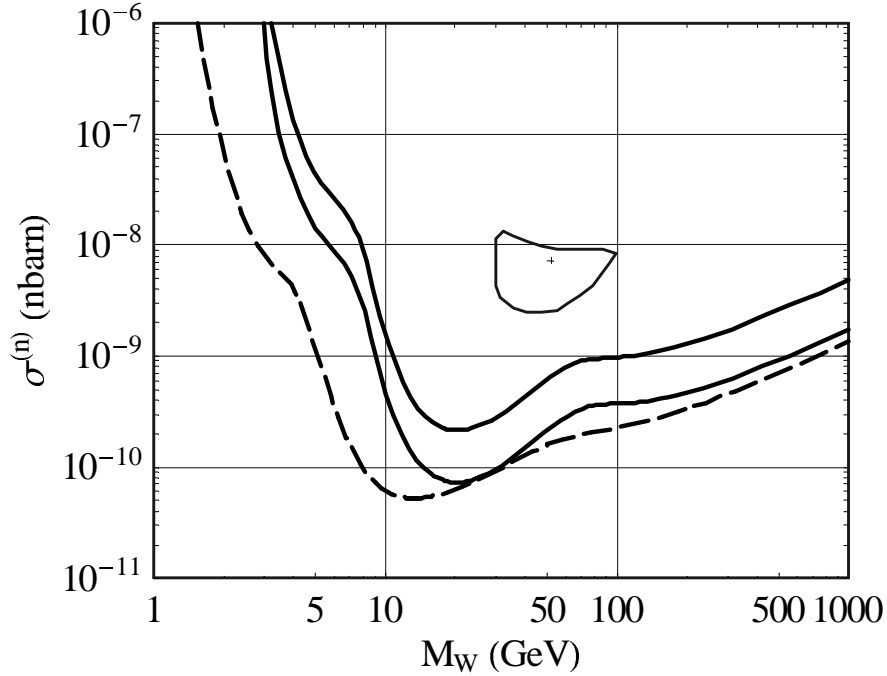,width=120mm} \caption{The
solid lines represent the sensitivity plot in the ($m,\sigma$)
plane for CUORE, assuming a threshold of 5 keV, two years of
exposure (1500 kg year) and flat backgrounds of 0.1 and 0.01
c/keV/kg/day. It has been calculated for $\langle\delta^2\rangle=$
5.6 (see the text). The sensitivity curve has been also calculated
for a possible threshold of 2 keV with a background of 0.01
c/keV/kg/day (dashed line). The closed contour represents the
3$\sigma$ CL region singled out by the modulation analysis
performed by the DAMA experiment \cite{Bernabei}, and the cross
indicates the minimum of the likelihood found by the same
authors.\label{cuore_mod}}
\end{center}
\end{figure}

Using this equation we have estimated the region that could be
within reach for CUORE and CUORICINO with the above mentioned
assumptions on the background levels. We have fixed a value of 5.6
for $\langle \delta^2 \rangle$ that corresponds to 50\%
probability of obtaining a positive result at 90\% C.L.. In figure
\ref{cuoricino_mod} it is shown the curves obtained for a
threshold of 5 keV, two years of exposure with CUORICINO (84 kg
year) and two assumed flat backgrounds of 1 and 0.1 c/keV/kg/day.
One can see that CUORICINO could already explore pretty well the
DAMA region looking for a positive annual modulation signal. In
figure \ref{cuore_mod} similar curves are presented, assuming flat
backgrounds of 0.1 and 0.01 c/keV/kg/day, two years of exposure of
CUORE (1500 kg year) and a threshold of 5 keV (solid lines). The
possibility of a lower thresholds of 2 keV with a background of
0.01 c/keV/kg/day is also shown (dashed line).

In conclusion, CUORE and CUORICINO will be able to explore and/or
exclude WIMPs lying in large regions of their parameter space. The
capability of CUORICINO / CUORE to investigate the DAMA region
through the exclusion plot (time integrated method) relies in
getting a background of 0.1 c/keV/kg/day from 5 keV onwards,
independently of more elaborated time modulation methods which
require an exhaustive control of the stability of the experiment.
However, CUORICINO and CUORE could also attempt to look for annual
modulation of WIMP signals provided that the stability of the
experiment be guaranteed.

\section{Solar axion detection}

Axions are light pseudoscalar particles which arise in theories in
which the Peccei-Quinn U(1) symmetry has been introduced to solve
the strong CP problem \cite{Peccei:1977hh}. They could have been
produced in early stages of the Universe being attractive
candidates for the cold dark matter (and in some particular
scenarios for the hot dark matter) responsible to 1/3 of the
ingredients of a flat universe. Dark matter axions can exist in
the mass window $10^{-2(3)}$ eV $<m_{a}\leq 10^{-6}$ eV, but
hadronic axions could exist with masses around the eV.

Axions could also be copiously produced in the core of the stars
by means of the Primakoff conversion of the plasma photons. In
particular, a nearby and powerful source of stellar axions would
be the Sun. The solar axion flux can be easily estimated
\cite{vanBibber:1989ge,Creswick} within the standard solar model,
resulting in an axion flux of an average energy of about 4 keV
that can produce detectable X-rays when reconverted again in an
electromagnetic field. Moreover, it has been pointed out recently
that the dimming of supernovae SNIa might be due to the conversion
of photons into axions in the extra-galactic magnetic field
\cite{Csaki}. That $\gamma$-a oscillation could make unobservable
about 1/3 of the SN emitted light and so, they would appear
fainter than implied by the luminosity-distance versus redshift
relation, without need to invoke an accelerated expansion of the
Universe. The SN result would be matched by axions of mass
$\sim10^{-16}$ eV and coupling to photons $\gagamma \sim2.5 \times
10^{-12}$ GeV$^{-1}$. So stellar axions may play an important role
in Cosmology. We would like to stress that, although we focus on
the axion because its special theoretical motivations, all this
scenario is also valid for a generic pseudoscalar (or scalar)
particle coupled to photons \cite{Masso:1995tw}. Needless to say
that the discovery of any type of pseudoscalar or scalar particle
would be extremely interesting in Particle Physics. We will keep
our discussion, however, restricted to the case of solar axions.

Crystal detectors provide a simple mechanism for solar axion
detection \cite{Paschos,Creswick}. Axions can pass in the
proximity of the atomic nuclei of the crystal where the intense
electric field can trigger their conversion into photons. The
detection rate is enhanced if axions from the Sun coherently
convert into photons when their incident angle with a given
crystalline plane fulfills the Bragg condition. This induces a
correlation of the signal with the position of the Sun which can
be searched for in the data and allows for background subtraction.
The potentiality of Primakoff conversion in crystals relies in the
fact that it can explore a range of axion masses ($m_a\lsim 0.1$
keV) not accessible to other direct searches. Moreover it is a
relatively simple technique that can be directly applied to
detectors searching for WIMPs.

Primakoff conversion using a crystal lattice has already been
employed in two germanium experiments: SOLAX \cite{SOLAX} and
COSME-II \cite{COSME} with the ensuing limits for axion-photon
coupling $\gagamma \lsim 2.7\times 10^{-9}$ GeV$^{-1}$ and
$\gagamma \lsim 2.8\times 10^{-9}$ GeV$^{-1}$ respectively. These
constraints are stronger than that of the Tokyo axion helioscope
\cite{tokyo} for $m_a \gsim 0.26$ eV and do not rely on
astrophysical considerations (i.e. on Red Giants or HB stars
dynamics \cite{Raffelt}). The orientation of the crystal was not
known so that the data were analyzed taking the angle
corresponding to the most conservative limit.

It has been noted that the model that yields the solar axion
fluxes used to calculate the expected signals is not compatible
with the constraints coming from helioseismology if $\gagamma
\gsim 10^{-9}$ GeV$^{-1}$ \cite{Schlattl}. This would imply a
possible inconsistency for solar axion limits above that value,
and sets a minimal goal for the sensitivity of future experiments.

The use of CUORE to search for solar axions via Bragg scattering
should have a priori some advantages with respect to germanium
detectors, because of the larger mass and the known orientation of
the crystals. On the other hand, as the cross-section for
Primakoff conversion depends on the square of the atomic number,
TeO$_{2}$ will be a priori a better candidate than Germanium.
Needless to say that a low energy threshold is mandatory because
the expected signal lies in the energy region 2 keV $\lsim E
\lsim$ 10 keV and is peaked at $E\simeq 4$ keV.
%Moreover, the strong time-variation of the signal
%requires a good resolution in order to be detected.

A detailed analysis has been performed \cite{Cebrian} for a
TeO$_2$ crystal (which has a tetragonal structure \cite{Crystal})
assuming different values for the experimental parameters. As it
is shown in Ref. \cite{Cebrian}, the bound on axion-photon
coupling which a given experiment can achieve can be estimated
through the expression:

$$ g_{a\gamma \gamma }<g_{a\gamma \gamma }^{\lim }\simeq k\left(
\frac b{\rm c/keV/kg/day}\frac{\rm kg}M\frac{\rm years}T\right)
^{1/8}\times 10^{-9}{\rm \ GeV}^{-1} $$

\noindent where $k$ depends on the crystal structure and material,
as well as on the experimental threshold and resolution. For the
case of TeO$_2$ and a threshold of 5 keV, $k$ has been calculated
to be $k=3.0$ assuming an energy resolution of 2 keV. The
computation of this expression for some assumed values of the
experimental parameters is shown in table \ref{tab_cuore} for
CUORICINO and CUORE. In all cases flat backgrounds and 2 years of
exposure are assumed.

\begin{figure}[h]
\begin{center}
\hspace{0 cm} \psfig{figure=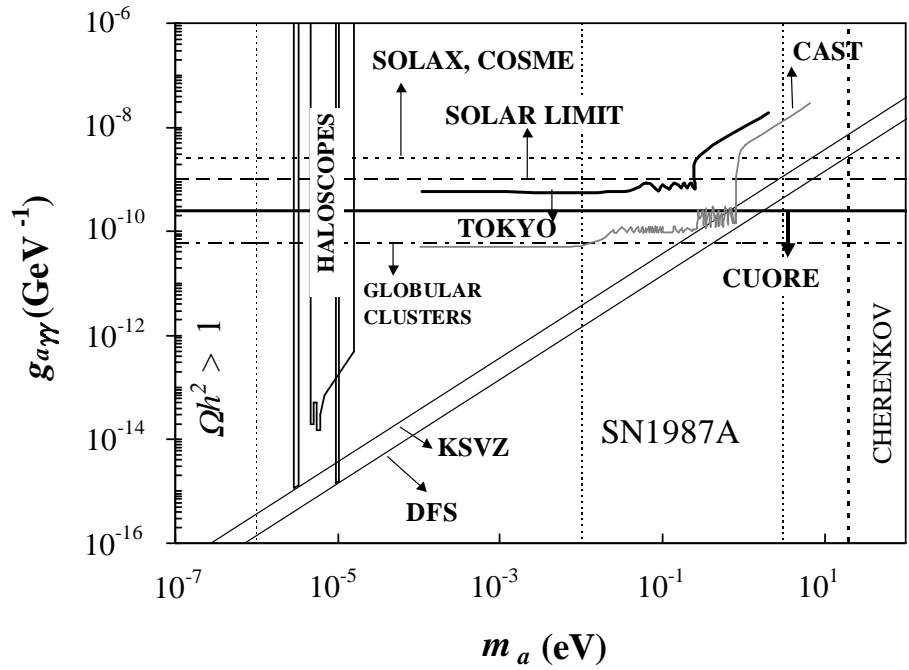,width=120mm}
\caption{Best bound attainable with CUORE (straight line labelled
"CUORE") compared with others limits.\label{fig}}
\end{center}
\end{figure}

It is worth to notice the faible dependence of the ultimate
achievable axion-photon coupling bound on the experimental
parameters, background and exposure MT: the 1/8 power dependence
of $g_{a\gamma\gamma}$ on such parameters softens their impact in
the final result. The best limit shown in table \ref{tab_cuore} is
in fact only one order of magnitude better than the present limits
of SOLAX and COSME-II. The $\gagamma$ bound that CUORE could
provide is depicted comparatively to other limits in figure
\ref{fig}.

The limit which can be expected from the CUORICINO experiment is
comparable to the helioseismological bound mentioned before. CUORE
could go even further (see Figure \ref{fig}). It should be
stressed that the bounds on $g_{a\gamma \gamma}$ obtained with
this technique stagnate at a few $\times 10^{-10}$ GeV$^{-1}$, not
too far from the goal expected for CUORE, as has been demonstrated
in \cite{Cebrian}. No bounds below these limits can be expected by
this technique from other crystal detectors like NaI, Ge or
TeO$_2$, and consequently, there are no realistic chances to
challenge the limit inferred from HB stars counting in globular
clusters \cite{Raffelt} and a discovery of the axion by CUORE
would presumably imply either a systematic effect in the
stellar-count observations in globular clusters or a substantial
change in the theoretical models that describe the late-stage
evolution of low-metallicity stars. To obtain lower values of
$\gagamma$ one should go to the magnet helioscopes like that of
Tokio \cite{tokyo} and that of CERN (CAST experiment \cite{cast}
currently being mounted). In particular, the best current
experimental bound of $\gagamma$ comes from the Tokyo helioscope:
$\gagamma \leq 6 \times 10^{-10}$ GeV$^{-1}$ for $m_{a} \lesssim
0.03$ eV and $\gagamma \leq 6.8-10.9 \times 10^{-10}$ GeV$^{-1}$
for $m_{a} \sim 0.05-0.27$ eV. The sensitivity of CAST is supposed
to provide a bound $\gagamma \leq 5 \times 10^{-11}$ GeV$^{-1}$ or
even lower.

Recently the DAMA collaboration has analyzed 53437 kg-day of data
of their NaI set up \cite{dama_axion}, in a search for solar
axions, following the techniques developed in ref. \cite{Cebrian},
where a calculation of the perspectives of various crystals
detectors (including NaI) for solar axion searches has been made.
The DAMA result $\gagamma \lsim 1.7\times 10^{-9}$ GeV$^{-1}$
improves slightly the limits obtained with other crystal detectors
\cite{SOLAX,COSME} and agrees with the result predicted in ref.
\cite{Cebrian}.

\section{Conclusions}

We have reported the perspectives of CUORE, a projected massive
760 kg array of 1000 TeO$_2$ bolometers, and of its first stage
CUORICINO, with 42 kg of the same crystals (currently being
mounted), as far as their physics potential to detect various
types of rare events is concerned. The estimated background and
resolution, based on Monte Carlo studies, together with the
results obtained in recent improvements in the performances of the
MIBETA experiment and the information obtained from the
preliminary tests of CUORICINO, have allowed us to assess the
potentialities of these experiments for double beta decay
searches, solar axion detection and WIMP exclusion or
identification. In these three types of searches, CUORE and to
some extent CUORICINO will be powerful tools to explore, with
higher sensitivity, such rare phenomena.

\section{Acknowledgements}

This work has been partially supported by the Spanish CICYT,
contract AEN99-1033 and the EU Network Contract
ERB-FMRX-CT-98-0167. We are indebted to the CUORE Collaboration
for allowing us to use unpublished results of the R\&D program of
CUORICINO and to E. Fiorini for his contribution to the conception
and coordination of the CUORE project. Useful discussion on the MC
simulation of the background with O. Cremonesi, M. Pavan and S.
Capelli are acknowledged. We also thank G. Luzon for discussions
on the cosmogenic activation of the crystals.

%\newpage

%\begin{table}[h] \centering
%\begin{tabular}{lcccccc} \hline\hline
%             &  Copper    & Archaeological  & Lead 1 & Lead 2& Tellurite & Mylar  \\
%             & (pg/g) & Lead (pg/g) & (Bq/kg) & (Bq/kg)& (pg/g) & (mBq/kg)  \\
%             \hline
%    $^{238}U$ & 1  & 1 &  &   &  0.01 & 9.5   \\

%    $^{232}Th$ &  10  & 10 &  &   &  0.01 & 26 \\

%    $^{40}K$ &  1  & 1 &  &   &  0.01 & 1480\\

%    $^{210}Pb$ &   & 4 mBq/kg & 160 &  16  &  &  \\
%           \hline\hline
%\end{tabular}
%\caption {Inputs for the assumed activities in different
%materials: copper, lead, tellurite and mylar.}\label{inputs}
%\end{table}

\newpage

%\begin{table}[tb] \centering
\begin{table}[h] \centering
  \begin{tabular}{cccc} \hline\hline
    $^{76}$Ge  &  $^{130}$Te &  $^{136}$Xe & nuclear model  \\ \hline
     $1.12\times 10^{-13}$ &  $5.33\times 10^{-13}$ & $1.18\times 10^{-13}$ &  QRPA \cite{QRPA1} \\
     $1.12\times 10^{-13}$ &  $4.84\times 10^{-13}$ & $1.87\times 10^{-13}$ &  QRPA \cite{QRPA2} \\
     $1.87\times 10^{-14}$ &  $3.96\times 10^{-13}$ & $7.9\times 10^{-14}$ &  QRPA \cite{QRPA3} \\
     $1.54\times 10^{-13}$ &  $1.63\times 10^{-12}$ &                        &  Weak Coupling SM \cite{WCSM} \\
     $1.13\times 10^{-13}$ &  $1.1\times 10^{-12}$ &                         &  Generalized Seniority \cite{GenSen} \\
     $1.21\times 10^{-13}$ &  $5.0\times 10^{-13}$ & $1.73\times 10^{-13}$ &  QRPA \cite{QRPA4} \\
     $7.33\times 10^{-14}$ &  $3.0\times 10^{-13}$ & $1.45\times 10^{-13}$ &  QRPA without pn pairing \cite{QRPA5} \\
     $1.42\times 10^{-14}$ &  $1.24\times 10^{-13}$ & $9.3\times 10^{-14}$ &  QRPA with pn pairing \cite{QRPA5} \\
     $5.8\times 10^{-13}$ &  $3.18\times 10^{-12}$ &                        &  \cite{Kla} \\
     $1.5\times 10^{-14}$ &                          & $2.16\times 10^{-14}$ &  Large basis SM \cite{LbSM} \\
     $9.5\times 10^{-14}$ &  $3.6\times 10^{-13}$ & $6.06\times 10^{-14}$ &  Operator expansion method \cite{OEM} \\

     \hline\hline
  \end{tabular}
  \caption{$2\beta 0 \nu$ nuclear merits $F_N^{0\nu}$ (y$^{-1}$)
  of emitters used in some source=detector calorimeters, according
  to various nuclear models.}\label{tab_doblebeta}
\end{table}

\newpage

\begin{table}[h] \centering
  \begin{tabular}{ccccc} \hline\hline
    {\bf Mass}  &  {\bf Resolution}&  {\bf Threshold}   & {\bf Background } &  {\bf $g_{a\gamma \gamma }^{\lim }$ (2 years)}\\
     {\bf (kg)}  &  {\bf (keV)}  &  {\bf (keV)} & {\bf (c/kg/keV/day)} &  {\bf (GeV$^{-1}$)}\\ \hline
     42       &  2   & 5    &  0.1 &  1.3$\times 10^{-9}$ \\
     760       &  2   & 5    &  0.01 &  6.7$\times 10^{-10}$ \\
%     760       &  2   & 5    &  0.005 &  6.2$\times 10^{-10}$ \\
%     760       &  2   & 5    &  0.001  &  5.0$\times 10^{-10}$\\
 %    760       &  1   & 5    &  0.01  &  6.5$\times 10^{-10}$\\
%     760       &  1   & 5    &  0.005 & 6.0$\times 10^{-10}$\\
%     760       &  1   & 5    &  0.001  &  4.9$\times 10^{-10}$\\
%     760       &  1   & 2    &  0.01  &  6.0$\times 10^{-10}$\\
%     760       &  1   & 2    &  0.005  &  5.6$\times 10^{-10}$\\
%     760       &  1   & 2    &  0.001  &  4.6$\times 10^{-10}$\\
     \hline\hline
  \end{tabular}
    \caption{Expected limits on the photon-axion coupling for 2 years of exposure of CUORICINO and CUORE assuming the quoted values
  for the experimental parameters}\label{tab_cuore}
\end{table}

%\newpage

%\begin{figure}[tb]
%\begin{figure}[h]
%\begin{center}
%\hspace{0 cm} \psfig{figure=simspclow.eps,width=120mm}
%\caption{Low energy region of the simulated intrinsic background
%spectra for CUORICINO before (dashed line) and after (solid line)
%the anticoincidence rejection.\label{simspclow}}
%\end{center}
%\end{figure}

%\newpage

%\begin{figure}[tb]
%\begin{figure}[h]
%\begin{center}
%\hspace{0 cm} \psfig{figure=simspchigh.eps,width=120mm}
%\caption{Simulated intrinsic background spectra for CUORICINO
%before (dashed line) and after (solid line) the anticoincidence
%rejection.\label{simspchigh}}
%\end{center}
%\end{figure}

\newpage

%\begin{figure}[tb]

\newpage

%\begin{figure}[tb]

\newpage

%\begin{figure}[tb]

\newpage

%\begin{figure}[tb]

\newpage

%\begin{figure}[tb]

\end{document}